\documentclass[aps,notitlepage,amsmath,amssymb]{revtex4}

\begin{document}

\title{Canonical Ensemble in Non-extensive Statistical Mechanics}

\author{Julius Ruseckas}

\email{julius.ruseckas@tfai.vu.lt}

\homepage{http://www.itpa.lt/~ruseckas}

\affiliation{Institute of Theoretical Physics and Astronomy, Vilnius University,
A.~Go\v{s}tauto 12, LT-01108 Vilnius, Lithuania}

\begin{abstract}
The framework of non-extensive statistical mechanics, proposed by
Tsallis, has been used to describe a variety of systems. The non-extensive
statistical mechanics is usually introduced in a formal way, using
the maximization of entropy. In this article we investigate the canonical
ensemble in the non-extensive statistical mechanics using a more traditional
way, by considering a small system interacting with a large reservoir
via short-range forces. The reservoir is characterized by generalized
entropy instead of the Boltzmann-Gibbs entropy. Assuming equal probabilities
for all available microstates we derive the equations of the non-extensive
statistical mechanics. Such a procedure can provide deeper insight
into applicability of the non-extensive statistics.
\end{abstract}

\maketitle

\section{Introduction}

Complexity in natural or artificial systems may be caused by long-range
interactions, long-range memory, non-ergodicity or multifractality.
Such systems have exotic thermodynamical properties and are unusual
from the point of view of traditional Boltzmann-Gibbs statistical
mechanics. Statistical description of complex systems can be provided
using the non-extensive statistical mechanics that generalizes the
Boltzmann-Gibbs statistics \cite{Tsallis2009-1,Tsallis2009-2,Telesca2010}.
The non-extensive statistical mechanics has been used to describe
phenomena in various in high-energy physics \cite{Adare2011}, spin-glasses
\cite{Pickup2009}, cold atoms in optical lattices \cite{Lutz2013},
trapped ions \cite{DeVoe2009}, anomalous diffusion \cite{Huang2010,Prehl2012},
dusty plasmas \cite{Liu2008}, low-dimensional dissipative and conservative
maps in the dynamical systems \cite{Afsar2013,Tirnakli2009,Ruiz2012},
turbulent flows \cite{Beck2013}, Langevin dynamics with fluctuating
temperature \cite{Budini2012,Du2012}. Concepts related to the non-extensive
statistical mechanics have found applications not only in physics
but in chemistry, biology, mathematics, economics, and informatics
as well \cite{Gell-Mann2004,Abe2006,Picoli2009}.

The basis of he non-extensive statistical mechanics is the generalized
entropy \cite{Tsallis2009-1} 
\begin{equation}
S_{q}=k_{\mathrm{B}}\frac{1-\sum_{\mu}p(\mu)^{q}}{q-1}\,,\label{eq:q-entr}
\end{equation}
where $p(\mu)$ is the probability of finding the system in the state
characterized by the parameters $\mu$; the parameter $q$ describes
the non-extensiveness of the system. More generalized entropies and
distribution functions are introduced in Refs.~\cite{Hanel2011-1,Hanel2011-2}.
The generalized entropy (\ref{eq:q-entr}) can be written in a form
similar to the Bolzmann-Gibbs entropy
\begin{equation}
S_{\mathrm{BG}}=-k_{\mathrm{B}}\sum_{\mu}p(\mu)\ln p(\mu)\label{eq:BG}
\end{equation}
as an average of $q$-logarithm \cite{Tsallis2009-1}:
\begin{equation}
S_{q}=k_{\mathrm{B}}\sum_{\mu}p(\mu)\ln_{q}\frac{1}{p(\mu)}\,,
\end{equation}
where the $q$-logarithm is defined as
\begin{equation}
\ln_{q}x=\frac{x^{1-q}-1}{1-q}\,.\label{eq:q-log1}
\end{equation}
In the limit $q\rightarrow1$ the $q$-logarithm becomes an ordinary
logarithm, thus the Boltzmann-Gibbs entropy can be obtained from Eq.~(\ref{eq:q-entr})
in the limit $q\rightarrow1$ \cite{Tsallis2009-1,Tsallis2009-2}.
The inverse function of the $q$-logarithm is the $q$-exponential
function
\begin{equation}
\exp_{q}(x)\equiv[1+(1-q)x]_{+}^{\frac{1}{1-q}}\,,\label{eq:q-exp1}
\end{equation}
with $[x]_{+}=x$ if $x>0$, and $[x]_{+}=0$ otherwise. The $q$-exponential
and $q$-logarithm appear in many equations of non-extensive statistical
mechanics \cite{Tsallis2009-1}. Some properties of $q$-exponential
and $q$-logarithm are presented in Appendix~\ref{sec:properties}.

The equilibrium of an isolated system consisting of $N$ particles
is described by the microcanonical ensemble. In the statistical physics
it is assumed that the equilibrium in the microcanonical ensemble
corresponds to equally probable microstates \cite{Landau1980,Landsberg2014},
therefore in the microcanonical ensemble $p(\mu)=1/W$, where $W$
is the number of microstates. Non-extensive statistical mechanics
can describe non-ergodic systems where not all available microstates
can be reached. In this case $W$ is the effective number of microstates,
that is the number of microstates whose probability is not zero. When
probabilities are equal, Eq.~(\ref{eq:q-entr}) for the generalized
entropy takes the simpler form 
\begin{equation}
S_{q}=k_{\mathrm{B}}\ln_{q}W\,.
\end{equation}
In the systems with long-range interactions and long-range correlations
the effective number of microstates $W$ can grow not exponentially
with the number of particles in the system $N$ but slower, as a power-law
of $N$. For such a systems the standard Boltzmann-Gibbs entropy (\ref{eq:BG})
is not proportional to the number of particles in the system and thus
is not extensive. The extensive quantity is the generalized entropy
(\ref{eq:q-entr}) for some value of $q\neq1$. In general, if the
entropy $S_{q}$ is proportional to the number of particles $N$ then
the number of microstates $W$ grows as $W\sim\exp_{q}N$. There are
two different cases: (i) $q<1$ and $W\sim N^{1/(1-q)}$. The number
of microstates grows as a power-law. (ii) $q>1$ and $W$ behaves
as $(1-(q-1)AN)^{-1/(q-1)}$. In this case there is a maximum value
of the number of particles $N_{\mathrm{crit}}$ where the number of
microstates becomes infinite and thus the macroscopic limit $N\rightarrow\infty$
cannot be taken. Because the of this complication occurring when $q>1$
in this paper we consider only the case of $q<1$; the value of $q$
in all the equations below should be assumed to be less than $1$.
The case of $q>1$ warrants a separate investigation and is outside
of the scope of the present paper.

The canonical ensemble in the non-extensive statistical mechanics
is usually introduced in a formal way, starting from the maximization
of the generalized entropy \cite{Tsallis2009-1}. The physical assumptions
appear in the maximization procedure in the form of constraints. However,
the $q$-averages used for constraints are unusual from the point
of view of ordinary, Boltzmann-Gibbs statistics. The physical justification
of $q$-averages and escort distributions is not completely clear.
Thus a more physically transparent method would be useful for understanding
the non-extensive statistical mechanics. The goal of this paper is
to investigate the canonical ensemble in the non-extensive statistical
mechanics using a more traditional way, by considering a small system
interacting with a large reservoir via short-range forces. Consistent
investigation of such a situation has not been performed yet. We assume
that the generalized entropy (\ref{eq:q-entr}) for some value of
$q<1$ instead of the Boltzmann-Gibbs entropy is extensive for the
reservoir. In addition, as in the standard statistical mechanics we
assume equal probabilities for all available microstates of the combined
system consisting of the small system and the reservoir. By doing
so we can avoid the critique of the generalized entropy presented
in Refs.~\cite{Wang2002,Neuenberg2003}.

The paper is organized as follows: To make the comparison of the non-extensive
statistical mechanics with the standard Boltzmann-Gibbs statistical
mechanics easier, in Section~\ref{sec:standard} we briefly present
the usual construction of the canonical ensemble in the standard statistical
mechanics. In Section~\ref{sec:non-extensive} we consider the canonical
ensemble in the non-extensive statistical mechanics and in Section~\ref{sec:Legendre}
we explore the resulting Legendre transformation structure. Section~\ref{sec:conclusions}
summarizes our findings.

\section{Canonical ensemble in Boltzmann\textendash Gibbs statistical mechanics}

\label{sec:standard}To highlight the differences from the non-extensive
statistical mechanics, let us at first briefly review the canonical
ensemble in the extensive Boltzmann-Gibbs statistical mechanics. The
standard approach \cite{Landau1980,Landsberg2014} is to consider
a composite system consisting of a system under investigation $\mathrm{S}$
interacting with a large reservoir $\mathrm{R}$. The system $\mathrm{S}$
has energy $E$, the energy of the reservoir $\mathrm{R}$ is $E_{\mathrm{R}}$
and the energy of the composite system is $E_{\mathrm{tot}}$. Due
to the interaction the system $\mathrm{S}$ and the reservoir $\mathrm{R}$
can exchange energy. The interaction is assumed to be short-range,
therefore in the macroscopic limit the energy of the interaction is
negligible and the total energy of the composite system is $E_{\mathrm{tot}}=E+E_{\mathrm{R}}$.
For simplicity we assume that there is no exchange of the particles
between the system $\mathrm{S}$ and the reservoir $\mathrm{R}$.

The number of microstates in the system $\mathrm{S}$ having the energy
$E$ is $W(E)$ and the number of microstates in the reservoir is
$W_{\mathrm{R}}(E_{\mathrm{R}})$. Here it is assumed that the numbers
of microstates depend only on the energy. The short range interactions
of the system $\mathrm{S}$ with the reservoir $\mathrm{R}$ do not
significantly change the numbers of microstates and thus the total
number of microstates in the combined system when the system $\mathrm{S}$
has energy $E$ is $W(E)W_{\mathrm{R}}(E_{\mathrm{tot}}-E)$. The
full number of microstates $W_{\mathrm{tot}}(E_{\mathrm{tot}})$ of
the combined system is obtained summing over all available energies
of the system $\mathrm{S}$:
\begin{equation}
W_{\mathrm{tot}}(E_{\mathrm{tot}})=\sum_{E}W(E)W_{\mathrm{R}}(E_{\mathrm{tot}}-E)\,.
\end{equation}
Introducing the entropy of the system $S(E)=k_{\mathrm{B}}\ln W(E)$
and the entropy of the reservoir $S_{\mathrm{R}}(E_{\mathrm{R}})=k_{\mathrm{B}}\ln W_{\mathrm{R}}(E_{\mathrm{R}})$
we can write
\begin{equation}
W_{\mathrm{tot}}(E_{\mathrm{tot}})=\sum_{E}e^{\frac{1}{k_{\mathrm{B}}}S(E)+\frac{1}{k_{\mathrm{B}}}S_{\mathrm{R}}(E_{\mathrm{tot}}-E)}\,.\label{eq:num-micro-1}
\end{equation}
In the extensive Boltzmann-Gibbs statistics the entropy of the reservoir
$S_{\mathrm{R}}$ is proportional to the number of particles $N_{\mathrm{R}}$
in the reservoir and is macroscopically large. The sum of large exponentials
can be approximated by the largest term, as is described in the Appendix~\ref{sec:sum-exponentials}.
In the statistical mechanics it is postulated that in the equilibrium
the probability of each microstate is the same and equal $1/W_{\mathrm{tot}}$.
Thus the most probable state of the composite system corresponds to
the largest term in the sum (\ref{eq:num-micro-1}). The most probable
energy $U$ of the system $\mathrm{S}$ corresponding to this largest
term can be found from the condition
\begin{equation}
\frac{\partial}{\partial U}S(U)-\frac{\partial}{\partial E_{\mathrm{tot}}}S_{\mathrm{R}}(E_{\mathrm{tot}}-U)=0\,.
\end{equation}
This condition allows to introduce the temperature $T$ characterizing
the equilibrium: 
\begin{equation}
\frac{1}{T}=\frac{\partial}{\partial U}S(U)=\frac{\partial}{\partial E_{\mathrm{tot}}}S_{\mathrm{R}}(E_{\mathrm{tot}}-U)\,.\label{eq:temp}
\end{equation}

We can also consider the situation when the Boltzmann-Gibbs entropy
of the system $\mathrm{S}$ is not necessarily extensive and proportional
to the number of particles $N$ in the system. If we introduce the
generalized entropy as $S_{q}(E)=k_{\mathrm{B}}\ln_{q}W(E)$ then
the sum (\ref{eq:num-micro-1}) becomes
\begin{equation}
W_{\mathrm{tot}}(E_{\mathrm{tot}})=\sum_{E}e_{q}^{\frac{1}{k_{\mathrm{B}}}S_{q}(E)}e^{\frac{1}{k_{\mathrm{B}}}S_{\mathrm{R}}(E_{\mathrm{tot}}-E)}
\end{equation}
and the largest term is determined from the condition
\begin{equation}
\left(e_{q}^{\frac{1}{k_{\mathrm{B}}}S_{q}(U)}\right)^{q-1}\frac{\partial}{\partial U}S_{q}(U)-\frac{\partial}{\partial E_{\mathrm{tot}}}S_{\mathrm{R}}(E_{\mathrm{tot}}-U)=0\,.
\end{equation}
Here we have used Eq.~(\ref{eq:q-exp-deriv}). We can conclude, that
the temperature $T$ is related to the generalized entropy $S_{q}$
via the equation
\begin{equation}
\frac{1}{T}=\frac{\frac{\partial}{\partial U}S_{q}(U)}{1+\frac{1-q}{k_{\mathrm{B}}}S_{q}(U)}\,.\label{eq:temp-non-ext}
\end{equation}
Introducing the auxiliary $q$-temperature by the equation
\begin{equation}
\frac{1}{T_{q}}=\frac{\partial}{\partial U}S_{q}(U)
\end{equation}
we get the relation
\begin{equation}
T=T_{q}\left(1+\frac{1-q}{k_{\mathrm{B}}}S_{q}(U)\right)\,.\label{eq:t-phys-t-q}
\end{equation}
The auxiliary temperature $T_{q}$ in the formulation of the non-extensive
statistical mechanics based on maximization of entropy can appear
as the inverse of the Lagrange multiplier associated with the energy
constraint. Althoug $T_{q}$ is not the physical temperature, it can
have another physical meaning. For example, such effective temperature
is directly related to the density of vortices in type II superconductors
\cite{Nobre2015}. The relation (\ref{eq:t-phys-t-q}) between the
physical temperature $T$ and the auxiliary temperature $T_{q}$ has
been proposed by various authors in Refs.~\cite{Abe1999,Rama2000,Abe2001a,Abe2001b,Abe2001,Martinez2001a,Martinez2001,Casas2002,Scarfone2010}.
Definitions of the temperature associated with different formulations
of the non-extensive statistical mechanics have been analyzed in Ref.~\cite{Wang2004}.
The general requirement that composition rules of entropy and energy
should satisfy to be compatible with zeroth law of thermodynamics
has been investigated in Ref.~\cite{Biro2011}. It has been shown
that formal logarithms of the original quantities should be additive.

Note, that the statistics of the system $\mathrm{S}$ is determined
by the reservoir, as we see from Eq.~(\ref{eq:prob-micro}). Therefore,
it is more convenient to describe even such a system using the Boltzmann-Gibbs
entropy. The same conclusion has been made in Ref.~\cite{Toral2003}:
it has been shown that physical temperature and pressure within the
formalism for non-extensive thermostatistics leads to expressions
which coincide with those obtained by using the standard formalism
of statistical mechanics.

According to the central postulate of the statistical mechanics, the
probability of the microstate $\mu\otimes\mu_{\mathrm{R}}$ where
the system $\mathrm{S}$ is in the microstate $\mu$ and the reservoir
is in the microstate $\mu_{\mathrm{R}}$is
\begin{equation}
p(\mu\otimes\mu_{\mathrm{R}})=\frac{1}{W(E_{\mathrm{tot}})}\,.
\end{equation}
The probability of the microstate $\mu$ of the system $\mathrm{S}$
then is
\begin{equation}
p(\mu)=\sum_{\mu_{\mathrm{R}}}p(\mu\otimes\mu_{\mathrm{R}})\,.
\end{equation}
If the energy of the microstate $\mu$ is $E_{\mu}$ then the energy
of the reservoir is $E_{\mathrm{tot}}-E_{\mu}$ and the number of
acceptable microstates of the reservoir is $W_{\mathrm{R}}(E_{\mathrm{tot}}-E_{\mu})$.
We obtain that the probability of the microstate is equal to
\begin{equation}
p(\mu)=\frac{W_{\mathrm{R}}(E_{\mathrm{tot}}-E_{\mu})}{W(E_{\mathrm{tot}})}\,.\label{eq:prob-micro}
\end{equation}
Approximating the number of microstates as
\begin{equation}
W_{\mathrm{R}}(E_{\mathrm{tot}}-E)=e^{\frac{1}{k_{\mathrm{B}}}S_{\mathrm{R}}(E_{\mathrm{tot}}-E)}\approx e^{\frac{1}{k_{\mathrm{B}}}S_{\mathrm{R}}(E_{\mathrm{tot}})-\frac{1}{k_{\mathrm{B}}}E\frac{\partial}{\partial E_{\mathrm{tot}}}S_{\mathrm{R}}(E_{\mathrm{tot}})}
\end{equation}
we obtain that the probability of the microstate of the system $\mathrm{S}$
is proportional to the Boltzmann factor
\begin{equation}
P(E)=\exp\left(-\frac{1}{k_{\mathrm{B}}T}E\right)\,.\label{eq:boltzmann}
\end{equation}
Here we used the definition of temperature (\ref{eq:temp}): $\frac{\partial}{\partial E_{\mathrm{tot}}}S_{\mathrm{R}}(E_{\mathrm{tot}})\approx\frac{1}{T}$.
Note, that for the justification of the exponential form of Eq.~(\ref{eq:boltzmann})
it is essential that the Bolzmann-Gibbs entropy of the large reservoir
had very small second derivative, $\frac{\partial^{2}}{\partial E_{\mathrm{tot}}^{2}}S_{\mathrm{R}}(E_{\mathrm{tot}})\approx0$.
This requirement means that the heat capacity of the reservoir
\begin{equation}
C_{\mathrm{R}}=-\frac{1}{T^{2}\frac{\partial^{2}}{\partial E_{\mathrm{tot}}^{2}}S_{\mathrm{R}}(E_{\mathrm{tot}})}\,.\label{eq:heat-cap}
\end{equation}
should be very large, that is the reservoir should be a thermostat.
Other possible forms of entropy (for example, the generalized entropy
$S_{q}^{(\mathrm{R})}=k_{\mathrm{B}}\ln_{q}W_{\mathrm{R}}$ with $q\neq1$)
do not have small second derivative and thus do not lead to a good
approximation for the probability of microstate.

From the Boltzmann factor (\ref{eq:boltzmann}) follows that the normalized
probability of the microstate can be written as
\begin{equation}
p(\mu)=\frac{1}{Z}e^{-\frac{1}{k_{\mathrm{B}}T}E_{\mu}}\,,
\end{equation}
where
\begin{equation}
Z=\sum_{\mu}e^{\frac{1}{k_{\mathrm{B}}T}E_{\mu}}
\end{equation}
is the partition function. The distribution of the energy of the system
$E$ is obtained multiplying the probability $p(\mu)$ by the number
of microstates having energy $E_{\mu}=E$. This number is equal to
$W(E)=e^{\frac{1}{k_{\mathrm{B}}}S(E)}$, therefore the distribution
of the energy is given by 
\begin{equation}
p(E)=\frac{1}{Z}e^{\frac{1}{k_{\mathrm{B}}}S(E)-\frac{1}{k_{\mathrm{B}}T}E}\,.
\end{equation}
The probability $p(E)$ should be normalized, thus the partition function
can be also written as
\begin{equation}
Z=\sum_{E}e^{\frac{1}{k_{\mathrm{B}}}S(E)-\frac{1}{k_{\mathrm{B}}T}E}\,.
\end{equation}
In the macroscopic limit the sum of large exponentials can be approximated
by the largest term, therefore 
\begin{equation}
\ln Z\approx\frac{1}{k_{\mathrm{B}}}S(U)-\frac{1}{k_{\mathrm{B}}T}U\,,\label{eq:appr-extensive}
\end{equation}
where the energy $U$ corresponding to the largest term is obtained
from the equation
\begin{equation}
\frac{\partial}{\partial U}S(U)=\frac{1}{T}\,.
\end{equation}

The average energy of the system
\begin{equation}
\bar{U}=\sum_{\mu}E_{\mu}p(\mu)
\end{equation}
can be determined form the partition function:
\begin{equation}
\bar{U}=k_{\mathrm{B}}T^{2}\frac{\partial}{\partial T}\ln Z\,.\label{eq:aver-energy}
\end{equation}
The free energy $F$ is introduced according to the equation 
\begin{equation}
F=-k_{\mathrm{B}}T\ln Z\,.\label{eq:free-energy}
\end{equation}
The equality
\begin{equation}
F\equiv\bar{U}-T\bar{S}\label{eq:free-energy-2}
\end{equation}
defines the average entropy $\bar{S}$. Combining Eqs.~(\ref{eq:aver-energy})--(\ref{eq:free-energy-2})
we get
\[
\frac{\partial\bar{S}}{\partial\bar{U}}=\frac{1}{T}\,.
\]
Due to approximation (\ref{eq:appr-extensive}) in the macroscopic
limit the average energy $\bar{U}$ coincides with the most probable
energy $U$ and the average entropy $\bar{S}$ coincides with $S(U)$.

\section{Canonical ensemble in non-extensive statistical mechanics}

\label{sec:non-extensive}Now let us consider the composite system
where the large reservoir $\mathrm{R}$ is not described by the Boltzmann-Gibbs
statistics. The Boltzmann-Gibbs entropy $S^{(\mathrm{R})}$ is not
proportional to the number of particles $N_{\mathrm{R}}$ in the reservoir
and is not extensive. The extensive quantity is the generalized entropy
$S_{q}^{(\mathrm{R})}$ for some value of $q\neq1$: $S_{q}^{(\mathrm{R})}\sim N_{\mathrm{R}}$.
Here we consider only the situation when $q<1$. The number of microstates
in such a reservoir is $W_{\mathrm{R}}=e_{q}^{\frac{1}{k_{\mathrm{B}}}S_{q}^{(\mathrm{R})}}\sim N_{\mathrm{R}}^{\frac{1}{1-q}}$.
Therefore, the Boltzmann-Gibbs entropy $S^{(\mathrm{R})}=k_{\mathrm{B}}\ln W_{\mathrm{R}}$
depends on the number of particles in the reservoir as $S_{\mathrm{R}}\sim\frac{1}{1-q}\ln N_{\mathrm{R}}$.
This expression for the Boltzmann-Gibbs entropy is similar to the
entropy of the system consisting of $d=\frac{1}{1-q}$ quasi-particles,
whereas the number of particles $N_{\mathrm{R}}$ plays the role of
the volume. A simple model of such a system has been presented in
Ref.~\cite{Ruseckas2015}: the model consist of a spin chain containing
$N_{\mathrm{R}}$ spins; spins next to each other have almost always
the same direction, except there are $d$ cases when the next spin
has an opposite direction. In the Boltzmann-Gibbs statistic large
reservoir has large heat capacity $C_{\mathrm{R}}$. Similarly, here
we require that the $q$-heat capacity of the reservoir, defined by
Eq.~(\ref{eq:q-heat-rez}), should be large.

The approach presented in this Section is similar to the approach
in Ref.~\cite{Abe2001}. However, in Ref.~\cite{Abe2001} the reservoir
is incorrectly interpreted as a heath bath and having large heat capacity.
As we have seen in the previous Section, such a reservoir leads to
the exponential Boltzmann factor and approximation of the expansion
of the number of states as a $q$-exponential is not justified. Interaction
of the system $\mathrm{S}$ weakly coupled to a finite reservoir having
a finite energy has been considered in Ref.~\cite{Plastino1994}.
Under the assumption that the number of microstates of the reservoir
having energy less than $E_{\mathrm{R}}$ grows as a power-law of
$E_{\mathrm{R}}$, the $q$-exponential distribution of the energy
of the system has been obtained. However, in Ref.~\cite{Plastino1994}
the extensivity of the generalized entropy has been not used and the
parameter $q$ tends to $1$ when the number of particles of the reservoir
increases.

As in the previous Section, the system under consideration $\mathrm{S}$
is interacting with the reservoir via short-range interactions, thus
the total energy of the composite system in the macroscopic limit
is $E_{\mathrm{tot}}=E+E_{\mathrm{R}}$ and the total number of microstates
in the composite system when the system $\mathrm{S}$ has energy $E$
is $W(E)W_{\mathrm{R}}(E_{\mathrm{tot}}-E)$. Introducing the generalized
entropy of the system $S_{q}(E)=k_{\mathrm{B}}\ln_{q}W(E)$ and the
generalized entropy of the reservoir $S_{q}^{(\mathrm{R})}(E_{\mathrm{R}})=k_{\mathrm{B}}\ln_{q}W_{\mathrm{R}}(E_{\mathrm{R}})$
we can write
\begin{equation}
W_{\mathrm{tot}}(E_{\mathrm{tot}})=\sum_{E}e_{q}^{\frac{1}{k_{\mathrm{B}}}S_{q}(E)}e_{q}^{\frac{1}{k_{\mathrm{B}}}S_{q}^{(\mathrm{R})}(E_{\mathrm{tot}}-E)}\,.\label{eq:num-micro-2}
\end{equation}
When $q<1$ this sum cannot be approximated by the largest term. Approximation
of a sum of large $q$-exponentials is investigated in Appendix~\ref{sec:sum-exponentials}.
We assume that the postulate of equal probabilities of microstates
in the equilibrium is valid also in the non-extensive statistical
mechanics. When the postulate of equal probabilities of microstates
is assumed, the statistics of the system $\mathrm{S}$ is determined
by the reservoir according to Eq.~(\ref{eq:prob-micro}). Therefore,
even an ordinary system interacting with the reservoir having large
$q$-heat capacity is more conveniently described by the $q$-entropy.

The largest term in the sum (\ref{eq:num-micro-2}) corresponds to
the most probable state of the composite system and is found from
the equation
\[
\frac{\frac{\partial}{\partial U}S_{q}(U)}{1+\frac{1-q}{k_{\mathrm{B}}}S_{q}(U)}-\frac{\frac{\partial}{\partial E_{\mathrm{tot}}}S_{q}^{(\mathrm{R})}(E_{\mathrm{tot}}-U)}{1+\frac{1-q}{k_{\mathrm{B}}}S_{q}^{(\mathrm{R})}(E_{\mathrm{tot}}-U)}=0\,.
\]
Here $U$ is the most-probable energy of the system $\mathrm{S}$.
Thus in order to satisfy the zeroth law of thermodynamics we need
to define the temperature $T$ as
\begin{equation}
\frac{1}{T}=\frac{\frac{\partial}{\partial U}S_{q}(U)}{1+\frac{1-q}{k_{\mathrm{B}}}S_{q}(U)}=\frac{\frac{\partial}{\partial E_{\mathrm{tot}}}S_{q}^{(\mathrm{R})}(E_{\mathrm{tot}}-U)}{1+\frac{1-q}{k_{\mathrm{B}}}S_{q}^{(\mathrm{R})}(E_{\mathrm{tot}}-U)}\,.\label{eq:temp-non-ext-2}
\end{equation}
This definition of the temperature is the same as Eq.~(\ref{eq:temp-non-ext}).

If one introduces the entropy of the combined system as $S_{q}^{(\mathrm{tot})}(E_{\mathrm{tot}})=k_{\mathrm{B}}\ln_{q}W(E_{\mathrm{tot}})$
then as a consequence of the impossibility to approximate the sum
(\ref{eq:num-micro-2}) by the largest term the entropy of the combined
system is not a simple combination of the entropies of the system
$\mathrm{S}$ and the reservoir $\mathrm{R}$: $S_{q}^{(\mathrm{tot})}(E_{\mathrm{tot}})\neq S_{q}(U)+S_{q}^{(\mathrm{R})}(E_{\mathrm{tot}}-U)+\frac{1-q}{k_{\mathrm{B}}}S_{q}(U)S_{q}^{(\mathrm{R})}(E_{\mathrm{tot}}-U)$.
Due to this the conclusions of Ref.~\cite{Ou2006} that the zeroth
law of thermodynamics holds only if the energy is also nonadditive
does not apply for the situation considered in this paper. On the
other hand, if one assumes that the interaction between the system
$\mathrm{S}$ and $\mathrm{R}$ are long range and the energy is not
additive then the pseudo-additivity of entropies can be valid \cite{Scarfone2010}.

The probability of a microstate of the system $\mathrm{S}$ is given
by Eq.~(\ref{eq:prob-micro}). Similarly as in the previous Section,
assuming that the second derivative of $q$-entropy of the reservoir
is very small, $\frac{\partial^{2}}{\partial E_{\mathrm{tot}}^{2}}S_{q}^{(\mathrm{R})}(E_{\mathrm{tot}})\approx0$,
we can approximate
\begin{equation}
W_{\mathrm{R}}(E_{\mathrm{tot}}-E)=e_{q}^{\frac{1}{k_{\mathrm{B}}}S_{q}^{(\mathrm{R})}(E_{\mathrm{tot}}-E)}\approx e_{q}^{\frac{1}{k_{\mathrm{B}}}S_{q}^{(\mathrm{R})}(E_{\mathrm{tot}}-U)-\frac{1}{k_{\mathrm{B}}}(E-U)\frac{\partial}{\partial E_{\mathrm{tot}}}S_{q}^{(\mathrm{R})}(E_{\mathrm{tot}}-U)}\,.\label{eq:q-expansion}
\end{equation}
Note, that now the Boltzmann-Gibbs entropy of the reservoir does not
have a small second derivative. The condition $\frac{\partial^{2}}{\partial E_{\mathrm{tot}}^{2}}S_{q}^{(\mathrm{R})}(E_{\mathrm{tot}})\approx0$
means that the $q$-heat capacity of the reservoir, defined in Eq.~(\ref{eq:q-heat-rez}),
is very large. The ordinary heat capacity of such a reservoir can
be determined as follows: if we increase the energy of the reservoir
by a small amount $\Delta E$, the increase of the temperature $T$,
according to Eq.~(\ref{eq:temp-non-ext-2}), is
\begin{equation}
\Delta T=\frac{1-q}{k_{\mathrm{B}}}\Delta E-T\frac{\frac{\partial^{2}}{\partial E_{\mathrm{R}}^{2}}S_{q}^{(\mathrm{R})}(E_{\mathrm{R}})}{\frac{\partial}{\partial E_{\mathrm{R}}}S_{q}^{(\mathrm{R})}(E_{\mathrm{R}})}\Delta E\,.
\end{equation}
This means that the heat capacity of the reservoir $C_{\mathrm{R}}=\frac{\Delta E}{\Delta T}$
is
\begin{equation}
C_{\mathrm{R}}=\frac{1}{\frac{1-q}{k_{\mathrm{B}}}+\frac{T}{T_{q}^{(R)}}\frac{1}{C_{q}^{(\mathrm{R})}}}\,,
\end{equation}
where $T_{q}^{(\mathrm{R})}$ is the auxiliary $q$-temperature of
the reservoir defined via the equation
\begin{equation}
\frac{1}{T_{q}^{(\mathrm{R})}}=\frac{\partial}{\partial E_{\mathrm{R}}}S_{q}^{(\mathrm{R})}(E_{\mathrm{R}})\label{eq:q-temp-rez}
\end{equation}
and
\begin{equation}
C_{q}^{(\mathrm{R})}=-\frac{1}{(T_{q}^{(\mathrm{R})})^{2}\frac{\partial^{2}}{\partial E_{\mathrm{R}}^{2}}S_{q}^{(\mathrm{R})}(E_{\mathrm{R}})}\label{eq:q-heat-rez}
\end{equation}
is the $q$-heat capacity of the reservoir, defined similarly to the
physical heat capacity, Eq.~(\ref{eq:heat-cap}). If the second derivative
of the generalized entropy is small,$\frac{\partial^{2}}{\partial E_{\mathrm{R}}^{2}}S_{q}^{(\mathrm{R})}(E_{\mathrm{R}})\approx0$,
then the heat capacity is $C=\frac{k_{\mathrm{B}}}{1-q}$. If we increase
the energy of the reservoir with very large $q$-heat capacity by
$\Delta E$, the new temperature of the reservoir becomes 
\begin{equation}
T^{\prime}=T+\frac{1-q}{k_{\mathrm{B}}}\Delta E\,.\label{eq:t-shift}
\end{equation}
The expression for the heat capacity $C=\frac{k_{\mathrm{B}}}{1-q}$
is the same as the heat capacity of a gas consisting of $d=\frac{1}{1-q}$
quasi-particles.

Using the property (\ref{eq:B-2}) of the $q$-exponential function
we obtain that the probability of the microstate of the system $\mathrm{S}$
is proportional to the factor
\begin{equation}
\tilde{P}(E)=\exp_{q}\left(-\frac{1}{k_{\mathrm{B}}T(U)}(E-U)\right)\,,
\end{equation}
where $T$ is the temperature according to Eq.~(\ref{eq:temp-non-ext-2}).
However, the temperature $T$ depends also on the properties of the
system, not only on the reservoir. It is more convenient to introduce
the temperature that the reservoir not interacting with the system
could have:
\begin{equation}
\frac{1}{T(0)}=\frac{\frac{\partial}{\partial E_{\mathrm{tot}}}S_{q}^{(\mathrm{R})}(E_{\mathrm{tot}})}{1+\frac{1-q}{k_{\mathrm{B}}}S_{q}^{(\mathrm{R})}(E_{\mathrm{tot}})}\,.
\end{equation}
Taking into account that $\frac{\partial^{2}}{\partial E_{\mathrm{R}}^{2}}S_{q}^{(\mathrm{R})}(E_{\mathrm{R}})\approx0$
we obtain
\begin{equation}
T(U)\approx\frac{1+\frac{1-q}{k_{\mathrm{B}}}S_{q}^{(\mathrm{R})}(E_{\mathrm{tot}})-\frac{1-q}{k_{\mathrm{B}}}U\frac{\partial}{\partial E_{\mathrm{tot}}}S_{q}^{(\mathrm{R})}(E_{\mathrm{tot}})}{\frac{\partial}{\partial E_{\mathrm{tot}}}S_{q}^{(\mathrm{R})}(E_{\mathrm{tot}})}=T(0)-\frac{1-q}{k_{\mathrm{B}}}U\,.\label{eq:tu-t0}
\end{equation}
This equation shows that the interaction with the system lowers the
temperature of the reservoir. On the other hand, the $q$-temperature
of the reservoir, defined by Eq.~(\ref{eq:q-temp-rez}) does not
change.

Using Eq.~(\ref{eq:tu-t0}) we get that the probability of the microstate
of the system $\mathrm{S}$ is proportional to
\begin{equation}
P(E)=\exp_{q}\left(-\frac{1}{k_{\mathrm{B}}T(0)}E\right)\,.\label{eq:factor-non-ext}
\end{equation}
An expression similar to Eq.~(\ref{eq:factor-non-ext}) has been
obtained in Ref.~\cite{Abe2001}. However, in Ref.~\cite{Abe2001}
the temperature that enters $P(E)$ has been interpreted as a physical
temperature $T$, because the reservoir has been assumed to be a thermostat.
The correct observation that the energy of the reservoir interacting
with the system should decrease has been presented in Ref.~\cite{Rama2000}.

One common objection to Eq.~(\ref{eq:factor-non-ext}) is that this
expression is not invariant to the change of zero of energies \cite{Tsallis2009-1}.
However, this reflects the physical situation of the system interacting
with the reservoir having very large $q$-heat capacity and, consequently,
small physical heat capacity. The zero of the energy of the system
$\mathrm{S}$ is fixed by the requirement that the energy of the reservoir
should be $E_{\mathrm{tot}}$ when $E=0$. If we shift the energy
zero by $\Delta E$, this is equivalent to the decrease of the energy
of the reservoir by $\Delta E$. This decrease of the energy of the
reservoir decreases the temperature. The probability of the microstate
should remain the same, thus the new factor should be proportional
to the old:
\begin{equation}
P^{\prime}(E)=\exp_{q}\left(-\frac{1}{k_{\mathrm{B}}T^{\prime}(0)}E\right)\sim P(E+\Delta E)=\exp_{q}\left(-\frac{1}{k_{\mathrm{B}}T(0)}(E+\Delta E)\right)\,.
\end{equation}
It follows that
\begin{equation}
T^{\prime}(0)=T(0)-\frac{1-q}{k_{\mathrm{B}}}\Delta E\,.
\end{equation}
This equation is consistent with Eq.~(\ref{eq:t-shift}). Similar
argument has been presented in Ref.~\cite{Plastino1994} by considering
a system $\mathrm{S}$ interacting with a finite reservoir.

Interesting feature of Eq.~(\ref{eq:factor-non-ext}) is the presence
of the cut-off energy: it follows from the definition of the $q$-exponential
function that $P(E)$ becomes zero when $E\geqslant E_{\mathrm{max}}$
where
\begin{equation}
E_{\mathrm{max}}=\frac{k_{\mathrm{B}}}{1-q}T(0)\,.
\end{equation}
This property of $P(E)$ ensures that the physical temperature $T$
is always positive. Discussion of possible cut-off prescriptions associated
with Tsallis' distributions is presented in Ref.~\cite{Teweldeberhan2005}.

Using the factor (\ref{eq:factor-non-ext}) we can write the normalized
probability of the microstate as
\begin{equation}
p(\mu)=\frac{1}{Z_{q}}e_{q}^{-\frac{1}{k_{\mathrm{B}}T(0)}E_{\mu}}\,,\label{eq:q-prob-normalized}
\end{equation}
where
\begin{equation}
Z_{q}=\sum_{\mu}e_{q}^{-\frac{1}{k_{\mathrm{B}}T(0)}E_{\mu}}
\end{equation}
is the generalized partition function. The distribution of the energy
of the system $E$ is obtained multiplying the probability $p(\mu)$
by the number $W(E)=e_{q}^{\frac{1}{k_{\mathrm{B}}}S_{q}(E)}$ of
microstates having energy $E_{\mu}=E$:
\begin{equation}
p(E)=\frac{1}{Z_{q}}e_{q}^{\frac{1}{k_{\mathrm{B}}}S_{q}(E)}e_{q}^{-\frac{1}{k_{\mathrm{B}}T(0)}E}=\frac{1}{Z_{q}}e_{q}^{\frac{1}{k_{\mathrm{B}}}\frac{T(E)}{T(0)}S_{q}(E)-\frac{1}{k_{\mathrm{B}}T(0)}E}\,,
\end{equation}
where
\begin{equation}
T(E)=T(0)-\frac{1-q}{k_{\mathrm{B}}}E\,.
\end{equation}
The probability $p(E)$ should be normalized, thus the partition function
can be also written as
\begin{equation}
Z_{q}=\sum_{E}e_{q}^{\frac{1}{k_{\mathrm{B}}}\frac{T(E)}{T(0)}S_{q}(E)-\frac{1}{k_{\mathrm{B}}T(0)}E}\,.\label{eq:zq-e}
\end{equation}
The energy $U$ corresponding to the largest term in the sum (\ref{eq:zq-e})
is determined by the equation
\begin{equation}
\frac{\frac{\partial}{\partial U}S_{q}(U)}{1+\frac{1-q}{k_{\mathrm{B}}}S_{q}(U)}=\frac{1}{T(U)}\,.
\end{equation}
According to Eq.~(\ref{eq:tu-t0}) the temperature $T(U)$ coincides
with the physical temperature.

\section{Generalized free energy}

\label{sec:Legendre}The probability proportional to the factor (\ref{eq:factor-non-ext})
admits several different possibilities to generalize the free energy.
First of all, there are three possibilities corresponding to three
temperatures: initial temperature of the reservoir $T(0)$, auxiliary
$q$-temperature $T_{q}$ and the physical temperature $T(U)$. From
those three choices only the temperature $T(0)$ depends only on the
reservoir and does not depend on the properties of the system. On
the other hand, the temperature $T(U)$ has a direct thermodynamical
interpretation. In addition, the average energy of the system is connected
to the generalized entropy with the parameter $2-q$. The derivative
of this generalized entropy with respect to average energy yields
another auxiliary temperature $T_{2-q}$ and the corresponding generalized
free energy.

\subsection{Initial temperature of the reservoir and unnormalized $q$-averages}

Let us consider first the generalized free energy $\bar{F}_{q}$ corresponding
to the temperature $T(0)$. This choice is closely related to the
approximation of the sum of large $q$-exponentials and to unnormalized
$q$-averages. When $q<1$ the sum in Eq.~(\ref{eq:zq-e}) cannot
be approximated by the largest term even in the macroscopic limit.
The approximate expression for the sum of large $q$-exponentials
is obtained in Appendix~\ref{sec:sum-exponentials}. According to
the results of Appendix~\ref{sec:sum-exponentials} and Eq.~(\ref{eq:zq-e})
the $q$-logarithm of $Z_{q}$ can be approximated as
\begin{equation}
\ln_{q}Z_{q}\approx\sum_{E}\left(\frac{1}{k_{\mathrm{B}}}S_{q}(E)-\frac{1+\frac{1-q}{k_{\mathrm{B}}}S_{q}(E_{\mu})}{k_{\mathrm{B}}T(0)}E\right)p(E)^{q}\,.\label{eq:lnqzq-appr}
\end{equation}
For any function of the energy $f(E)$ the following equality holds:
\begin{equation}
\sum_{E}f(E)p(E)^{q}=\sum_{E_{\mu}}f(E_{\mu})e_{q}^{\frac{1}{k_{\mathrm{B}}}S_{q}(E_{\mu})}\left(e_{q}^{\frac{1}{k_{\mathrm{B}}}S_{q}(E_{\mu})}\right)^{q-1}p(\mu)^{q}=\sum_{\mu}\frac{f(E_{\mu})}{1+\frac{1-q}{k_{\mathrm{B}}}S_{q}(E_{\mu})}p(\mu)^{q}\,.\label{eq:sum-e-sum-mu}
\end{equation}
Therefore, we can approximate the $q$-logarithm of $Z_{q}$ as
\begin{equation}
\ln_{q}Z_{q}\approx\sum_{\mu}\left(\frac{\frac{1}{k_{\mathrm{B}}}S_{q}(E_{\mu})}{1+\frac{1-q}{k_{\mathrm{B}}}S_{q}(E_{\mu})}-\frac{1}{k_{\mathrm{B}}T(0)}E_{\mu}\right)p(\mu)^{q}\,.
\end{equation}
This equation suggest to introduce the unnormalized $q$-average energy
of the system
\begin{equation}
\bar{U}_{q}=\sum_{\mu}E_{\mu}p(\mu)^{q}=\sum_{E}E\left(1+\frac{1-q}{k_{\mathrm{B}}}S_{q}(E)\right)p(E)^{q}\,.
\end{equation}
This unnormalized $q$-average of the energy can be determined from
the generalized partition function $Z_{q}$ using the equation
\begin{equation}
\bar{U}_{q}=k_{\mathrm{B}}T(0)^{2}\frac{\partial}{\partial T(0)}\ln_{q}Z_{q}\,.\label{eq:u1}
\end{equation}
In analogy to Eq.~(\ref{eq:free-energy}) we introduce the generalized
free energy
\begin{equation}
\bar{F}_{q}=-k_{\mathrm{B}}T(0)\ln_{q}Z_{q}\,.\label{eq:fq1}
\end{equation}
The equation 
\begin{equation}
\bar{F}_{q}\equiv\bar{U}_{q}-T(0)\bar{S}_{q}\label{eq:fq1-2}
\end{equation}
defines the entropy $\bar{S}_{q}$ which is related to the unnormalized
$q$-average of the entropy $S_{q}$. Using Eqs.~(\ref{eq:u1})--(\ref{eq:fq1-2})
we obtain
\begin{equation}
\frac{\partial\bar{S}_{q}}{\partial\bar{U}_{q}}=\frac{1}{T(0)}\,.\label{eq:q-entr-deriv}
\end{equation}

Entropy $\bar{S}_{q}$ can be calculated using the probabilities $p(\mu)$
according to Eq.~(\ref{eq:q-entr}). Indeed, we have
\begin{equation}
\bar{S}_{q}=\frac{1}{T(0)}(\bar{U}_{q}-\bar{F}_{q})=\sum_{\mu}\frac{1}{T(0)}E_{\mu}p(\mu)^{q}+k_{\mathrm{B}}\ln_{q}Z_{q}\,.\label{eq:tmp-1}
\end{equation}
Expressing the energy from the probability $p(\mu)$ we get
\begin{equation}
E_{\mu}=-k_{\mathrm{B}}T(0)\ln_{q}[p(\mu)Z_{q}]=-k_{\mathrm{B}}T(0)\left(\ln_{q}p(\mu)+p(\mu)^{1-q}\ln_{q}Z_{q}\right)\,.\label{eq:e-mu-expr}
\end{equation}
Inserting this expression for the energy $E_{\mu}$ into Eq.~(\ref{eq:tmp-1})
and taking into account that $\sum_{\mu}p(\mu)=1$ we obtain
\begin{equation}
\bar{S}_{q}=k_{\mathrm{B}}\frac{\sum_{\mu}p(\mu)^{q}-1}{1-q}\,.\label{eq:q-entr-2}
\end{equation}
This expression is consistent with the approximation (\ref{eq:lnqzq-appr}).
According to the approximation (\ref{eq:lnqzq-appr}) the entropy
$\bar{S}_{q}$ is
\begin{equation}
\bar{S}_{q}\approx\sum_{E}S_{q}(E)p(E)^{q}=\sum_{\mu}\frac{S_{q}(E_{\mu})}{1+\frac{1-q}{k_{\mathrm{B}}}S_{q}(E_{\mu})}p(\mu)^{q}\,.
\end{equation}
In the macroscopic limit the entropy $S_{q}(E)$ is large and we can
approximate 
\begin{equation}
\bar{S}_{q}\approx k_{\mathrm{B}}\sum_{\mu}\frac{p(\mu)^{q}}{1-q}\approx k_{\mathrm{B}}\frac{\sum_{\mu}p(\mu)^{q}-1}{1-q}\,.
\end{equation}
This expression is the same as (\ref{eq:q-entr-2}).

\subsection{$q$-temperature and normalized $q$-averages}

It can be more convenient to deal with normalized $q$-averages. The
normalized $q$-average of the energy is
\begin{equation}
U_{q}=\frac{\sum_{\mu}E_{\mu}p(\mu)^{q}}{\sum_{\mu}p(\mu)^{q}}\,.
\end{equation}
Using Eq.~(\ref{eq:sum-e-sum-mu}) the normalized $q$-average of
the energy can be written as
\begin{equation}
U_{q}=\frac{\sum_{E}E\left(1+\frac{1-q}{k_{\mathrm{B}}}S_{q}(E)\right)p(E)^{q}}{\sum_{E}\left(1+\frac{1-q}{k_{\mathrm{B}}}S_{q}(E)\right)p(E)^{q}}
\end{equation}
The sums of the form $\sum_{E}f(E)p(E)^{q}$ can be approximated as
$f(U)\sum_{E}p(E)^{q}+f^{\prime}(U)\sum_{E}(E-U)p(E)^{q}$. The sum
$\sum_{E}(E-U)p(E)^{q}$ is small, since close to the maximum $U$
the probability $p(E)$ is an even function of $E-U$. We obtain that
in the macroscopic limit the normalized $q$-average of energy $U_{q}$
should be close to the most probable energy $U$. Since the sum of
large $q$-exponentials cannot be approximated by the largest term
when $q<1$, it is not possible to determine the most probable energy
$U$ or the average energy $\sum_{\mu}E_{\mu}p(\mu)$ knowing only
the generalized partition function $Z_{q}$. However, it is possible
to calculate $U_{q}$, which is close to $U$. On the other hand,
the entropy $\bar{S}_{q}$ cannot be approximated by $S_{q}(U)$.

Using Eq.~(\ref{eq:q-entr-2}) the normalized $q$-average $U_{q}$
can be related to the unnormalized $\bar{U}_{q}$ via the equation
\begin{equation}
U_{q}=\frac{\bar{U}_{q}}{1+\frac{1-q}{k_{\mathrm{B}}}\bar{S}_{q}}\,.\label{eq:uq-norm-unnorm}
\end{equation}
From Eq.~(\ref{eq:uq-norm-unnorm}) follows that the introduction
of the normalized $q$-average energy $U_{q}$ allows to factorize
the generalized partition function $Z_{q}$:
\begin{equation}
Z_{q}=e_{q}^{\frac{1}{k_{\mathrm{B}}}\bar{S}_{q}-\frac{1}{k_{\mathrm{B}}T(0)}\bar{U}_{q}}=e_{q}^{\frac{1}{k_{\mathrm{B}}}\bar{S}_{q}}e_{q}^{-\frac{1}{k_{\mathrm{B}}T(0)}U_{q}}\,.\label{eq:zq-fact}
\end{equation}
The entropy of the combined system $S_{q}^{(\mathrm{tot})}$ using
the expansion (\ref{eq:q-expansion}) can be written as
\begin{equation}
S_{q}^{(\mathrm{tot})}=k_{B}\ln_{q}e_{q}^{\frac{1}{k_{\mathrm{B}}}S_{q}^{(\mathrm{R})}(E_{\mathrm{tot}}-U)}e_{q}^{\frac{1}{k_{\mathrm{B}}T(U)}U}Z_{q}
\end{equation}
Using Eq.~(\ref{eq:zq-fact}) and assuming that $U\approx U_{q}$
we get that the entropy $S_{q}^{(\mathrm{tot})}$ can be expressed
as the usual pseudo-additive combination of entropies from the non-extensive
statistical mechanics: 
\begin{equation}
S_{q}^{(\mathrm{tot})}\approx S_{q}^{(\mathrm{R})}(E_{\mathrm{tot}}-U_{q})+\bar{S}_{q}+\frac{1-q}{k_{B}}S_{q}^{(\mathrm{R})}(E_{\mathrm{tot}}-U_{q})\bar{S}_{q}\,.
\end{equation}
However, in this equation the entropy $\bar{S}_{q}$ is not directly
connected to the number of microstates of the system $\mathrm{S}$.

Let us introduce an auxiliary $q$-temperature $T_{q}$ of the system
$\mathrm{S}$ via the equation
\begin{equation}
\frac{1}{T_{q}}=\frac{\partial\bar{S}_{q}}{\partial U_{q}}\,.\label{eq:tq-tilde}
\end{equation}
Using Eqs.~(\ref{eq:q-entr-deriv}) and (\ref{eq:uq-norm-unnorm})
we get
\begin{equation}
T(U_{q})=T_{q}\left(1+\frac{1-q}{k_{\mathrm{B}}}\bar{S}_{q}\right)\,,\label{eq:tq-tilde-t}
\end{equation}
where $T(U_{q})=T(0)-\frac{1-q}{k_{\mathrm{B}}}U_{q}$ is the temperature
of the reservoir corresponding to the energy of the system equal to
$U_{q}$. Since $\bar{S}_{q}>0$, the $q$-temperature is always smaller
than the physical temperature $T(U_{q})$. Note, that only physical
temperatures of the system and the reservoir are equal. The $q$-temperature
of the system $T_{q}$ is not equal to the $q$-temperature of the
reservoir $T_{q}^{(\mathrm{R})}$.

We introduce the $q$-analog of the free energy corresponding to the
temperature $T_{q}$:
\begin{equation}
F_{q}\equiv U_{q}-T_{q}\bar{S}_{q}\,.\label{eq:fq-tilde}
\end{equation}
Then, using Eqs.~(\ref{eq:tq-tilde}) and (\ref{eq:fq-tilde}) we
get
\begin{equation}
\bar{S}_{q}=-\frac{\partial F_{q}}{\partial T_{q}}\,.
\end{equation}
We define the $q$-heat capacity of the system as
\begin{equation}
C_{q}=\frac{\partial U_{q}}{\partial T_{q}}=T_{q}\frac{\partial\bar{S}_{q}}{\partial T_{q}}=-T_{q}\frac{\partial^{2}F_{q}}{\partial T_{q}^{2}}\,.
\end{equation}
The physical heat capacity $C$ can be determined as the derivative
of $U_{q}$ with respect to the physical temperature $T(U_{q})$:
\begin{equation}
C=\frac{\partial U_{q}}{\partial T(U_{q})}\,.
\end{equation}
Using Eqs.~(\ref{eq:tq-tilde}) and (\ref{eq:tq-tilde-t}) we get
the equation that relates the physical heat capacity with the auxiliary
$q$-heat capacity: 
\begin{equation}
C=\frac{1}{\frac{T(U_{q})}{T_{q}}\frac{1}{C_{q}}+\frac{1-q}{k_{\mathrm{B}}}}\,.\label{eq:c-cq-tilde}
\end{equation}
Since $T_{q}<T(U_{q})$, from Eq.~(\ref{eq:c-cq-tilde}) follows
that the physical heat capacity $C$ is always smaller than the $q$-heat
capacity $C_{q}$.

Generalized partition function $\bar{Z}_{q}$ related to the generalized
free energy $F_{q}$ is
\begin{equation}
\bar{Z}_{q}\equiv e_{q}^{-\frac{F_{q}}{k_{\mathrm{B}}T_{q}}}\,.\label{eq:zq-tilde}
\end{equation}
Note that $Z_{q}\neq\bar{Z}_{q}$. Using Eqs.~(\ref{eq:tq-tilde}),
(\ref{eq:fq-tilde}) and (\ref{eq:zq-tilde}) we get the expression
for the energy $U_{q}$:
\begin{equation}
U_{q}=k_{\mathrm{B}}T_{q}^{2}\frac{\partial}{\partial T_{q}}\ln_{q}\bar{Z}_{q}\,.
\end{equation}
The generalized partition function $\bar{Z}_{q}$ cannot be directly
expressed as a sum. However, $\bar{Z}_{q}$ can be connected to a
sum of $q$-exponentials as follows: we write the probability of the
microstate in the form
\begin{equation}
p(\mu)=\frac{1}{\tilde{Z}_{q}}e_{q}^{-\frac{1}{k_{\mathrm{B}}T(U_{q})}(E_{\mu}-U_{q})}\,,
\end{equation}
where
\begin{equation}
\tilde{Z}_{q}=\sum_{\mu}e_{q}^{-\frac{1}{k_{\mathrm{B}}T(U_{q})}(E_{\mu}-U_{q})}
\end{equation}
is related to $Z_{q}$ via the equation
\begin{equation}
\tilde{Z}_{q}=Z_{q}e_{q}^{\frac{1}{k_{\mathrm{B}}T(U_{q})}U_{q}}\,.
\end{equation}
Using Eq.~(\ref{eq:zq-fact}) we get
\begin{equation}
\tilde{Z}_{q}=e_{q}^{\frac{1}{k_{\mathrm{B}}}\bar{S}_{q}}\,.
\end{equation}
Therefore,
\begin{equation}
\ln_{q}\bar{Z}_{q}=\ln_{q}\tilde{Z}_{q}-\frac{1}{k_{\mathrm{B}}T_{q}}U_{q}\,.
\end{equation}

\subsection{Physical temperature and R\'enyi entropy}

The third possibility is to in introduce the free energy corresponding
to the physical temperature $T(U_{q})$. In order to do this let us
consider another entropy, given by the equation
\begin{equation}
\tilde{S}_{q}=k_{\mathrm{B}}\ln e_{q}^{\frac{1}{k_{\mathrm{B}}}\bar{S}_{q}}=\frac{k_{\mathrm{B}}}{1-q}\ln\left(1+\frac{1-q}{k_{\mathrm{B}}}\bar{S}_{q}\right)\,.
\end{equation}
The entropy $\tilde{S}_{q}$ is more directly connected to the physical
temperature $T(U_{q})$. Indeed, using Eqs.~(\ref{eq:tq-tilde})
and (\ref{eq:tq-tilde-t}) we get that the derivative of the entropy
$\tilde{S}_{q}$ gives the physical temperature:
\begin{equation}
\frac{\partial\tilde{S}_{q}}{\partial U_{q}}=\frac{1}{T(U_{q})}\,.\label{eq:renyi-prop}
\end{equation}
From Eq.~(\ref{eq:q-entr-2}) it follows that
\begin{equation}
\tilde{S}_{q}=\frac{k_{\mathrm{B}}}{1-q}\ln\left(\sum_{\mu}p(\mu)^{q}\right)\,.
\end{equation}
Thus the entropy $\tilde{S}_{q}$ is the R\'enyi entropy \cite{Renyi1970,Tsallis1988}.

Using the R\'enyi entropy we introduce the free energy corresponding
to the physical temperature $T(U_{q})$:
\begin{equation}
\tilde{F}_{q}\equiv U_{q}-T(U_{q})\tilde{S}_{q}\,.
\end{equation}
Also in this case we retain the Legendre transformation structure.
For example, using Eq.~(\ref{eq:renyi-prop}) we get
\begin{equation}
\frac{\partial\tilde{F}_{q}}{\partial T(U_{q})}=-\tilde{S}_{q}\,.
\end{equation}

\subsection{Average energy of the system}

It is impossible to exactly determine the average energy of the system
\begin{equation}
\bar{U}=\sum_{\mu}E_{\mu}p(\mu)\label{eq:u-aver}
\end{equation}
knowing only the sum of $q$-exponents $Z_{q}$. However, the knowledge
of another sum 
\[
\sum_{\mu}\left(e_{q}^{-\frac{1}{k_{\mathrm{B}}T(0)}E_{\mu}}\right)^{2-q}
\]
allows us to do so. Indeed, using the property of the $q$-exponential
function (\ref{eq:q-exp-deriv}) and the expression for the probability
$p(\mu)$ Eq.~(\ref{eq:q-prob-normalized}) we get
\begin{equation}
\bar{U}=\frac{k_{\mathrm{B}}T(0)^{2}}{(2-q)Z_{q}}\frac{\partial}{\partial T(0)}\sum_{\mu}\left(e_{q}^{-\frac{1}{k_{\mathrm{B}}T(0)}E_{\mu}}\right)^{2-q}\,.
\end{equation}
Instead of this sum we can use the generalized entropy (\ref{eq:q-entr})
with the parameter $q^{\prime}=2-q$:
\begin{equation}
\bar{S}_{2-q}=k_{\mathrm{B}}\frac{1-\sum_{\mu}p(\mu)^{2-q}}{1-q}\,.\label{eq:entr-2-q}
\end{equation}
Using the generalized entropy $\bar{S}_{2-q}$ the expression for
the average energy $\bar{U}$ becomes
\begin{equation}
\bar{U}=\frac{k_{\mathrm{B}}T(0)^{2}}{(2-q)Z_{q}}\frac{\partial}{\partial T(0)}Z_{q}^{2-q}\left(1-\frac{1-q}{k_{\mathrm{B}}}\bar{S}_{2-q}\right)\,.\label{eq:u-aver-1}
\end{equation}
We can obtain another expression for the average energy $\bar{U}$
by inserting $E_{\mu}$ from Eq.~(\ref{eq:e-mu-expr}) into Eq.~(\ref{eq:u-aver}):
\begin{equation}
\bar{U}=T(0)Z_{q}^{1-q}\bar{S}_{2-q}-k_{\mathrm{B}}T(0)\ln_{q}Z_{q}\,.\label{eq:u-aver-2}
\end{equation}
Combining Eqs.~(\ref{eq:u-aver-1}) and (\ref{eq:u-aver-2}) we obtain
\begin{equation}
\frac{\partial\bar{U}}{\partial T(0)}-\frac{T(0)Z_{q}^{1-q}}{2-q}\frac{\partial\bar{S}_{2-q}}{\partial T(0)}=0\,.
\end{equation}
Therefore, yet another auxiliary temperature $T_{2-q}$, introduced
by the equation
\begin{equation}
\frac{1}{T_{2-q}}=\frac{\partial\bar{S}_{2-q}}{\partial\bar{U}}\,,
\end{equation}
is equal to
\begin{equation}
T_{2-q}=T(0)\frac{Z_{q}^{1-q}}{2-q}\,.
\end{equation}
This relation between temperatures is exactly the same as obtained
by maximizing the entropy (\ref{eq:entr-2-q}) with the constraint
(\ref{eq:u-aver}) \cite{Tsallis2009-1}. The temperature $T_{2-q}$
, similarly as the temperature $T_{q}$, depends not only on the reservoir
but also on the properties of the system.

The generalized free energy corresponding to the average internal
energy of the system $\bar{U}$ and the temperature $T_{2-q}$ is
\begin{equation}
F_{2-q}=\bar{U}-T_{2-q}\bar{S}_{2-q}\,.
\end{equation}
This expression for the generalized free energy $F_{2-q}$ is similar
to the expression 
\begin{equation}
\bar{F}_{q}=\bar{U}-(2-q)T_{2-q}\bar{S}_{2-q}
\end{equation}
for the generalized free energy $\bar{F}_{q}$ that follows from Eq.~(\ref{eq:u-aver-2}).
We see that in general $\bar{F}_{q}\neq F_{2-q}$ .

\section{Discussion}

\label{sec:conclusions} In summary, we have demonstrated that a small
system interacting with a large reservoir having large $q$-heat capacity
can be described by the non-extensive statistical mechanics. From
the point of view of the ordinary statistics such a reservoir is similar
to a gas of $d=1/(1-q)$ quasi-particles. The probability of the microstate
of the system interacting with the reservoir via short-range forces
is given by the $q$-exponential function (\ref{eq:factor-non-ext}),
instead of the exponential Boltzmann factor (\ref{eq:boltzmann}).
Large $q$-heat capacity of the reservoir leads to a small physical
heat capacity, therefore the temperature in the equilibrium $T$ depends
both on the properties of the reservoir and the properties of the
system. In order to avoid this inconvenience one can consider the
temperature $T(0)$ of the reservoir that is not interacting with
the system or introduce an auxiliary $q$-temperature $T_{q}^{(\mathrm{R})}$
that remains constant due to large $q$-heat capacity of the reservoir.
Small heat capacity of the reservoir does not allow to consider it
as a thermostat, thus the description using the standard canonical
ensemble of the statistical mechanics is not applicable. The treatment
of the canonical ensemble presented in this paper allows us to obtain
relations between the physical temperature $T$ and the auxiliary
$q$-temperature $T_{q}$ (\ref{eq:tq-tilde-t}) as well as between
the $q$-heat capacity $C_{q}$ and the physical heat capacity $C$
(\ref{eq:c-cq-tilde}).

Sums of large exponentials often appear in the Boltzmann-Gibbs statistical
mechanics. Such sums can be approximated by keeping only the largest
term. Similarly, in the non-extensive statistical mechanics appear
sums of large $q$-exponentials. However, for such sums taking only
the largest term is a very poor approximation. This is because the
$q$-exponential function with $q<1$ does not decrease as fast as
the exponential function. As a consequence, the deviations from the
most probable state in the non-extensive statistical mechanics are
much larger than the deviations in the standard statistical mechanics.
As it is shown in Appendix~\ref{sec:properties}, sums of large $q$-exponentials
are well approximated using $q$-averages. This fact is one the reasons
why $q$-averages play such an important role in the non-extensive
statistics.

In this paper we considered the reservoir for which the generalized
entropy with $q<1$ is extensive. The case of $q>1$ is more complicated,
because one cannot take the macroscopic limit $N\rightarrow\infty$.
The investigation of the small system interacting with the reservoir
characterized by $q>1$ remains a task for the future.

\appendix

\section{Sum of large $q$-exponentials}

\label{sec:sum-exponentials}One can easily show that the sum of large
exponentials
\begin{equation}
Z=\sum_{i=1}^{W}e^{N\phi(i)}
\end{equation}
can be approximated by the largest term. Indeed, if $\phi_{\mathrm{max}}$
is the maximum of $\phi(i)$ then
\begin{equation}
e^{N\phi_{\mathrm{max}}}\leqslant Z\leqslant We^{N\phi_{\mathrm{max}}}
\end{equation}
and
\begin{equation}
0\leqslant\frac{\ln Z}{N}-\phi_{\mathrm{max}}\leqslant\frac{\ln W}{N}\,.
\end{equation}
If $W$ grows slower than exponentially with increasing $N$, then
in the limit of large $N$ the ratio $\ln W/N$ vanishes and we have
\begin{equation}
\lim_{N\rightarrow\infty}\frac{\ln Z}{N}=\phi_{\mathrm{max}}\,.
\end{equation}

Now let us consider the sum of large $q$-exponentials
\begin{equation}
Z_{q}=\sum_{i=1}^{W}e_{q}^{N\phi(i)}\,,
\end{equation}
where $q<1$. In contrast to the sum of large exponentials, approximation
of the sum of large $q$-exponentials with the largest term is a poor
one. We can construct a better approximation as follows: let us introduce
the weights
\begin{equation}
p(i)=\frac{e_{q}^{N\phi(i)}}{Z_{q}}
\end{equation}
and the unnormalized $q$-average
\begin{equation}
\langle\phi\rangle_{q}\equiv\sum_{i=1}^{W}\phi(i)p(i)^{q}\,.
\end{equation}
By noticing that 
\begin{equation}
N\phi(i)=\ln_{q}[p(i)Z_{q}]=\ln_{q}p(i)+p(i)^{1-q}\ln_{q}Z_{q}
\end{equation}
one can write the difference $\ln_{q}Z_{q}-N\langle\phi\rangle_{q}$
as
\begin{equation}
\ln_{q}Z_{q}-N\langle\phi\rangle_{q}=\frac{\sum_{i=1}^{W}p(i)^{q}-1}{1-q}\,.
\end{equation}
The sum $\sum_{i}p(i)^{q}$ can have the largest possible value when
all weights $p(i)$ are equal. In such a case $p(i)=1/W$ and $\sum_{i}p(i)^{q}=W^{1-q}$.
Thus
\begin{equation}
\frac{\ln_{q}Z_{q}}{N}-\langle\phi\rangle_{q}\leqslant\frac{\ln_{q}W}{N}\,.
\end{equation}
If $W$ grows with increasing $N$ as $N^{p}$ and
\begin{equation}
p<\frac{1}{1-q}
\end{equation}
then in the limit of large $N$ the ratio $\ln_{q}W/N$ vanishes and
we have
\begin{equation}
\lim_{N\rightarrow\infty}\frac{\ln_{q}Z_{q}}{N}=\langle\phi\rangle_{q}\,.
\end{equation}
This gives the required approximation of the sum of $q$-exponentials.
Note, that $\ln_{q}W/N$ as the upper limit of the difference $\ln_{q}Z_{q}-N\langle\phi\rangle_{q}$
is the worst case, when all terms in the sum are equal. For sufficiently
fast decreasing terms the sum $\sum_{i}p(i)^{q}$ can be bounded even
for larger $W$.

\section{Some properties of $q$-exponential function}

\label{sec:properties}In this paper we have used the following properties
of $q$-exponential and $q$-logarithm: multiplication of two $q$-exponentials
\begin{eqnarray}
e_{q}^{x}e_{q}^{y} & = & e_{q}^{[1+(1-q)y]x+y}=e_{q}^{x+[1+(1-q)x]y}\,,\label{eq:B-1}\\
e_{q}^{x+y} & = & e_{q}^{x}e_{q}^{\frac{y}{1+(1-q)x}}=e_{q}^{y}e_{q}^{\frac{x}{1+(1-q)y}}\,,\label{eq:B-2}
\end{eqnarray}
$q$-logarithm of a product
\begin{equation}
\ln_{q}xy=[1+(1-q)\ln_{q}y]\ln_{q}x+\ln_{q}y=\ln_{q}x+[1+(1-q)\ln_{q}x]\ln_{q}y\,,
\end{equation}
the derivatives of $q$-exponential and $q$-logarithm:
\begin{eqnarray}
\frac{d}{dx}e_{q}^{x} & = & (e_{q}^{x})^{q}\,,\label{eq:q-exp-deriv}\\
\frac{d}{dx}\ln_{q}x & = & \frac{1}{x^{q}}\,.\label{eq:B-5}
\end{eqnarray}
The equations (\ref{eq:B-1})--(\ref{eq:B-5}) can be easily derived
using the definitions (\ref{eq:q-exp1}) and are presented in the
Appendix~A of Ref.~\cite{Tsallis2009-1}.

\end{document}